\documentclass[aps,prl, reprint,superscriptaddress]{revtex4-1}  
\usepackage{graphicx}  
\usepackage{dcolumn}   
\usepackage{bm}        
\usepackage{amssymb}   
\usepackage{epstopdf}

\begin{document}

\widetext


\title{Measurements of individual and ensemble lifetimes of high-lying Rb Rydberg states}


\author{M. Archimi}
\affiliation{Dipartimento di Fisica ``E. Fermi'', Universit\`a di Pisa, Largo Bruno Pontecorvo 3, 56127 Pisa, Italy}

\author{C. Simonelli}
\affiliation{Dipartimento di Fisica ``E. Fermi'', Universit\`a di Pisa, Largo Bruno Pontecorvo 3, 56127 Pisa, Italy}
\affiliation{INO-CNR, Via G. Moruzzi 1, 56124 Pisa, Italy}

\author{L. Di Virgilio}
\affiliation{Dipartimento di Fisica ``E. Fermi'', Universit\`a di Pisa, Largo Bruno Pontecorvo 3, 56127 Pisa, Italy}

\author{A. Greco}
\affiliation{Dipartimento di Fisica ``E. Fermi'', Universit\`a di Pisa, Largo Bruno Pontecorvo 3, 56127 Pisa, Italy}

\author{M. Ceccanti}
\affiliation{Dipartimento di Fisica ``E. Fermi'', Universit\`a di Pisa, Largo Bruno Pontecorvo 3, 56127 Pisa, Italy}

\author{E. Arimondo}
\affiliation{Dipartimento di Fisica ``E. Fermi'', Universit\`a di Pisa, Largo Bruno Pontecorvo 3, 56127 Pisa, Italy}
\affiliation{INO-CNR, Via G. Moruzzi 1, 56124 Pisa, Italy}

\author{D. Ciampini}
\affiliation{Dipartimento di Fisica ``E. Fermi'', Universit\`a di Pisa, Largo Bruno Pontecorvo 3, 56127 Pisa, Italy}
\affiliation{INO-CNR, Via G. Moruzzi 1, 56124 Pisa, Italy}

\author{I.I. Ryabtsev}
\affiliation{Rzhanov Institute of Semiconductor Physics SB RAS, 630090 Novosibirsk State University, Russia }
\affiliation{Novosibirsk State University, 630090 Novosibirsk, Russia}

\author{I.I. Beterov}
\affiliation{Rzhanov Institute of Semiconductor Physics SB RAS, 630090 Novosibirsk State University, Russia }
\affiliation{Novosibirsk State University, 630090 Novosibirsk, Russia}
\affiliation{Novosibirsk Technical University, 630072 Novosibirsk, Russia}

\author{O. Morsch.}
\affiliation{Dipartimento di Fisica ``E. Fermi'', Universit\`a di Pisa, Largo Bruno Pontecorvo 3, 56127 Pisa, Italy}
\affiliation{INO-CNR, Via G. Moruzzi 1, 56124 Pisa, Italy}

\date{\today}

\begin{abstract}
We demonstrate a hybrid method based on field ionization and state-selective de-excitation capable of measuring the lifetimes of high-lying Rydberg states. For nS Rydberg states of Rb atoms with principal quantum number $60\leq n\leq88$, we measure both the individual target state lifetimes and those of the ensemble of Rydberg states populated via black-body radiation-induced transitions. We find good overall agreement with numerical calculations of the expected lifetimes in both cases. However, for the target state lifetimes, we find a local deviation towards shorter lifetimes for states around $n=72$, which we interpret as a signature of a modified black-body spectrum in the finite volume in which our experiments take place. 
\end{abstract}

\pacs{34.20.Cf, 32.80.Ee}
\maketitle

Rydberg states of atoms are characterised, amongst other properties \cite{gallagherxx,loew12}, by long radiative lifetimes. Compared to low-lying excited atomic states, which typically decay on timescales of less than a microsecond, high-lying Rydberg states (with principal quantum number above $n\approx 50$) can live for hundreds of microseconds. This makes them attractive for applications, {\em e.g}, in quantum simulation and quantum computation \cite{saffman10}, where the ground state and a Rydberg state can be used to encode quantum bits. In those applications, it is important also to take into account another peculiarity of high-lying Rydberg states, which is their interaction with black-body radiation \cite{gallagher79,farley81,cooke80,beterov09, beterovNJP}. The resulting transition rates to other nearby Rydberg states can be comparable to or even larger than those due to spontaneous decay to low-lying states. This creates practical problems when measuring Rydberg state lifetimes, as the different Rydberg states populated by black-body radiation are close in energy and, therefore, can be difficult to distinguish experimentally. 

So far, a number of studies have addressed the issue of Rydberg state lifetimes  \cite{gallagher75,kocher77,chupka93,merkt94,deoliveira02,feng09,branden10,mack15}. Experimentally, different techniques such as field ionization and state-selective field ionization \cite {gallagher76a,nosbaum95,hollenstein01}, trap loss spectroscopy  \cite{day08}, monitor states  \cite{branden10}, and all-optical methods based on probe beam absorption have been used \cite{mack15}. For alkali atoms such as rubidium, which we use in the present study, lifetimes have been measured up to principal quantum numbers around $n=45$  \cite{branden10}, and good agreement with numerical calculations has been found. For higher $n$, the typically employed methods become increasingly difficult to apply. State-selective field ionization for high-lying states above $n\approx60$ requires one to distinguish states whose ionization thresholds differ only by a few percent (for instance, between the 80S and 81S states, it changes by $5\%$ from $9.18\,\mathrm{V/cm}$ to $8.72\,\mathrm{V/cm}$, and the values for the nearest n’P states are even closer). Other techniques may also be unable to reliably measure the population of an individual Rydberg state, and instead only give information on a range of closely spaced Rydberg levels.\\

Here, we demonstrate a hybrid field ionization and state-selective laser de-excitation method that allows us to measure, in principle, Rydberg state lifetimes for almost arbitrary principal quantum numbers (limited essentially by the linewidth of the de-excitation laser and by residual electric fields) as well as the lifetimes of the corresponding “ensembles”, {\em i.e.}, of all Rydberg states that are populated by black-body radiation starting from the initially excited state, called the “target state”. Our measurements for nS Rydberg states with principal quantum numbers between 60 and 88 show good overall agreement with numerical calculations for both lifetimes. However, for the target state lifetime, we find a local deviation from theory towards shorter lifetimes around $n=72$, which we discuss in the context of the black-body spectrum inside the finite volume of our vacuum cell.\\

In our experiments, we excite nS Rydberg states of 87-Rb atoms in a standard magneto-optical trap (MOT) containing around 200,000 atoms at temperature $T\approx150\,\mathrm{\mu K}$ in a roughly spherical cloud $300\,\mathrm{\mu m}$ in size (Gaussian width; for details of the apparatus see  \cite{viteau10,viteau11}). The Rydberg states are reached from the $\rm 5S_{1/2}(F=2)$ ground state via the intermediate $\rm 6P_{3/2}(F=3)$ state using two co-propagating laser beams at $420\,\mathrm{nm}$ and $1013\,\mathrm{nm}$, with blue detuning $\Delta/2\pi = 37\,\mathrm{MHz}$ of the $420\,\mathrm{nm}$  laser from the $\rm 5S_{1/2}(F=2) -  6P_{3/2}(F=3)$ resonance (in each experimental cycle, the MOT beams are switched off for $1500\,\mathrm{\mu s}$ a few hundred ns before the excitation pulse, while the magnetic field remains switched on). Rydberg atoms are then detected by field ionization, in which a high-voltage pulse is applied to two pairs of electrodes placed outside the glass vacuum cell and the resulting ions are accelerated towards a channel electron multiplier. The overall detection efficiency is around $40\%$, and the highest electric fields achievable in our apparatus correspond to the field ionization threshold for Rb Rydberg states with $n \approx 60$. We have checked that above that threshold the detection efficiency is largely independent of $n$.\\

 \begin{figure}[htbp]
\begin{center}
\includegraphics[width=8cm]{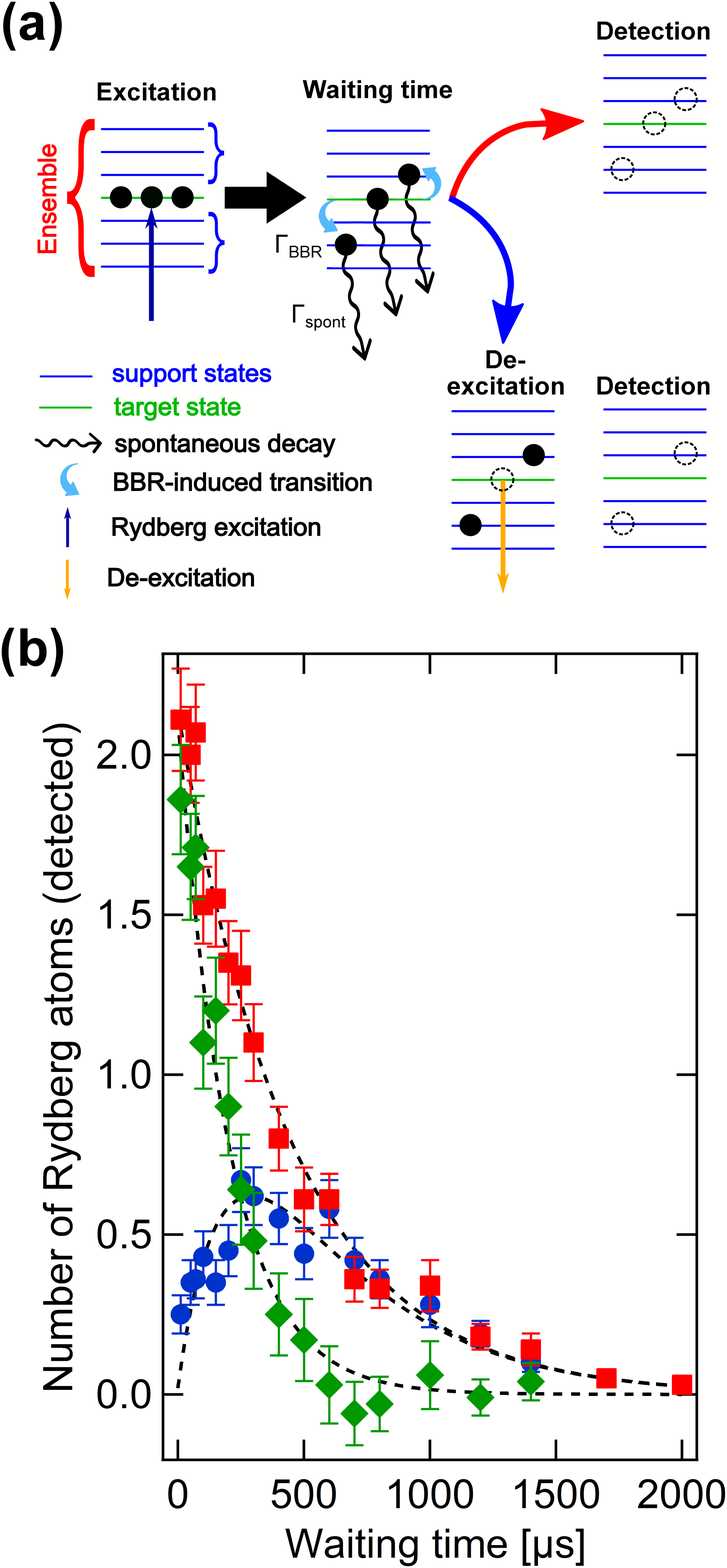}
\caption{Measurement of Rydberg target state and ensemble lifetimes. (a) Schematic of the measurements. After excitation of the target state, the number of excitations in the ensemble is measured by simple field ionization. The number of excitations in the support states, by contrast, is measured by first depumping the target state atoms to low-lying states using a $5\,\mathrm{\mu m}$ laser pulse resonant with the $\rm 6P_{3/2}(F=3)$ state, and then field ionizing the remaining excitations. (b) A typical measurement of the number of Rydberg excitations in the ensemble (red squares) and support states (blue circles) as a function of time for a $80S$ target state. The number of excitations in the target state (green diamonds) is then calculated from the difference of the ensemble and support states. The dashed lines are exponential fits, from which the target state and ensemble lifetimes are extracted (for the support data, the difference of the exponential fits for the ensemble and target was calculated). Error bars are one standard error of the mean.}
\label{default}
\end{center}
\end{figure}

A combination of field ionization and state-selective de-excitation  \cite{simonelli17} now gives us access to two quantities, as shown schematically in Fig. 1(a). After laser excitation of an initial Rydberg target state, a first measurement consists in waiting a variable time $t$ after the excitation pulse (typical duration around 300 ns) and then applying the electric field ionization pulse. During the waiting time two processes take place: spontaneous decay of atoms in the target state to low-lying states (around $95\%$ of the atoms decay to states with $n$ between 5 and 17, calculated using \cite{sibalic17}), and absorption as well as stimulated emission of photons of the black-body radiation inside the glass cell, which populates nearby Rydberg states with principal quantum number $n^\prime$ and angular momentum L (from which, in turn, other nearby states can be populated in a multi-step process \cite{beterovNJP}). While atoms that have undergone spontaneous decay can no longer be field ionized, those that have exchanged photons with the black-body radiation typically end up in states that are detected as Rydberg states in field ionization. We now fit an exponential decay to the measured number of Rydberg atoms $N_{\rm ens}(t)$ (see Fig. 1 (b)) and call the resulting $1/e$-time $\tau_{\rm ens}$ the ensemble lifetime. Given the practical constraints, the 'ensemble' denotes all Rydberg $n^\prime$L states that can be field ionized in our experiment. \\

The second measurement is similar to the first one, except that before field ionization, a $5\,\mathrm{\mu s}$ de-excitation laser pulse with Rabi frequency $\Omega/2\pi$ between 2 and $4\,\mathrm{MHz}$ resonantly couples the target Rydberg state to the $\rm 6P_{3/2}(F=3)$  intermediate state. This is achieved by changing the frequency of a double-pass acousto-optic modulator in the beam path of the $1013\,\mathrm{nm}$ laser, whereby the frequency of the de-excitation pulse can be suddenly (within $100\,\mathrm{ns}$) increased by $37\,\mathrm{MHz}$, matching the detuning $\Delta/2\pi$ of the excitation pulse. The  $\rm 6P_{3/2}$ state has a lifetime of $120\,\mathrm{ns}$, so effectively the de-excitation laser pulse depumps target state atoms to the ground state. After the de-excitation pulse, field ionization as described above measures the number of Rydberg atoms left in the cloud. In the case of a $100\%$ depumping efficiency $\alpha$, only atoms that have been re-distributed out of the target state to nearby Rydberg states, which are not resonant with the de-excitation pulse, will contribute to the field ionization signal (in practice, $\alpha$ is around $95\%$, and the duration of the pulse is chosen to be much shorter than the expected target state lifetime, whilst still resulting in a reasonable depumping efficiency). We call these states the 'support' of the  target state, and a typical measurement of the support population $N_{\rm supp}(t)$ is shown in Fig. 1 (b). \\

From  $N_{\rm ens}(t)$  and $N_{\rm supp}(t)$ we can now infer the number of atoms in the target Rydberg state $N_{\rm tar}(t) $ from the relations $N(t)=N_{\rm ens}(t)=N_{\rm tar}(t)+N_{\rm supp}(t)$ (before depumping) and $N^\prime(t)=(1-\alpha)N_{\rm tar}(t)+N_{\rm supp}(t)$ (after depumping). Fitting an exponential decay to $N_{\rm tar}(t)$, we obtain $\tau_{\rm tar}$. A typical measurement of $\tau_{\rm tar}$ for the $80S$ state is shown in Fig. 1(b).  We note that for the target state lifetimes reported in this work, the free fall and expansion of the atomic cloud between the excitation and depumping pulses (which leads to a decrease in the overlap between the spatial distribution of Rydberg atoms and the de-excitation beam) is not a substantial limitation, but could be relevant for lifetimes exceeding $1000\,\mathrm{\mu s}$ (we discuss this in more detail later).  \\

In principle, with our technique it is possible to measure target state populations as long as the closest S or D states (which can be coupled to the 6P state by the de-excitation laser) are separated in energy from the target state by more than the linewidth of the de-excitation laser (around $500\,\mathrm{kHz}$ in our apparatus, including residual Doppler broadening). In practice, however, there are further limitations. First, van der Waals interactions (which are responsible for the well-known dipole blockade effect \cite{browaeys, saffman1}) or dipole-dipole interactions between Rydberg atoms (which can arise, for instance, between $n$S and $n^\prime$P states) can shift the target state energy levels, resulting in an effective detuning of the de-excitation laser and hence a variation in the depumping efficiency $α\alpha$ over time, which will lead to systematic errors in the inferred lifetime. In order to limit such effects, we adjusted the two-photon Rabi frequency and duration of the excitation pulse such as to excite only $\approx 4$  Rydberg atoms on average (corrected for the detection efficiency) in a cloud defined by the overlap between the MOT and the two excitation lasers with waists $40\,\mathrm{\mu m}$ ($420\,\mathrm{nm}$ laser) and $90\,\mathrm{\mu m}$ ($1013\,\mathrm{nm}$ laser). The mean spacing between two Rydberg atoms (assuming a uniform distribution inside the cloud) was, therefore, on the order of $50\,\mathrm{\mu m}$, and hence van der Waals and dipole-dipole interactions could, to first approximation, be neglected (for instance, the van der Waals interaction between two 70S atoms at that distance is less than $100\,\mathrm{Hz}$). \\

A second limitation of our technique stems from electric fields in the vacuum cell, to which Rydberg atoms are extremely sensitive due to their large polarizability, which scales roughly as $n^7$. From measurements of the Stark map (target $n$S state and the $(n-3)$ Stark manifold) and comparison with numerical calculations \cite{sibalic17}, we determined the background electric field in our cell to be $215\pm10\,\mathrm{mV/cm}$ (the error includes day-to-day fluctuations). This field is likely due to rubidium atoms adsorbed on the inner surface of the cell, charged dust particles on the outside of the cell, and other field sources whose origin we are unable to establish with certainty. By applying a few tens of volts to the field ionization electrodes we were able to compensate a part of the background field, leading to a minimum residual field of around $100\,\mathrm{mV/cm}$. 

 \begin{figure}[htbp]
\begin{center}
\includegraphics[width=9cm]{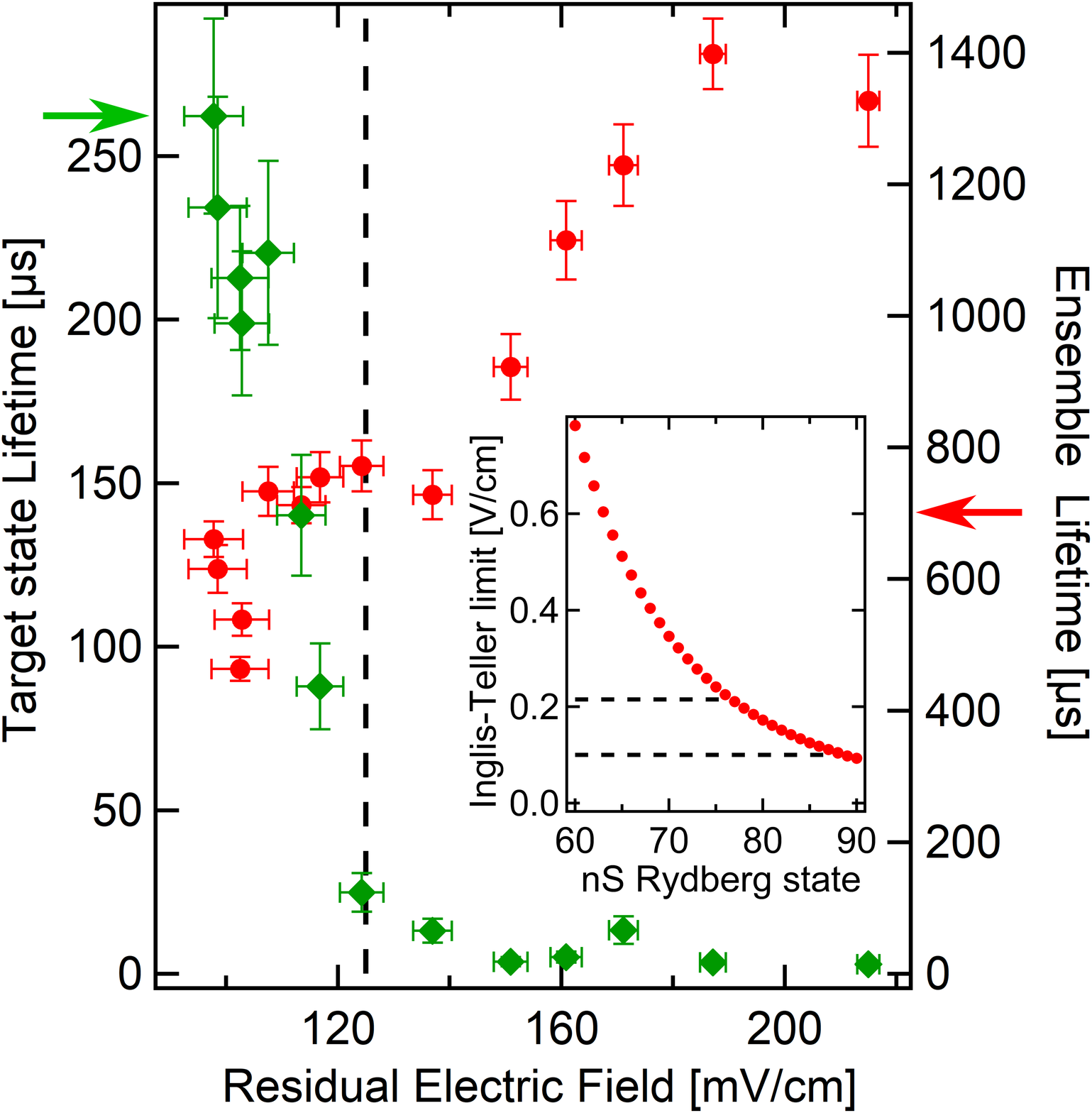}
\caption{Effect of the residual electric field $E$ in the cell on the measured target state (green diamonds) and ensemble lifetimes (red circles). Below the Inglis-Teller limit for the $85S$ state used here (vertical line), both lifetimes agree with the theoretical predictions (arrows) of  $258\,\mathrm{\mu s}$ for the target state and $719\,\mathrm{\mu s}$ for the ensemble lifetime. Error bars are one standard error of the mean (for the electric field values, we estimate an error based on our measurement protocol). Inset:  The Inglis-Teller limit for the range of principal quantum numbers considered in this work (calculated using \cite{sibalic17}). The dashed horizontal lines indicate the electric field in the cell with (bottom) and without compensation (top). }
\label{default}
\end{center}
\end{figure}

To study the effect of a finite electric field $E$ on our lifetime measurements, we measured the target state and ensemble lifetimes as a function of $E$ (Fig. 2). In particular, we scanned the electric field around the Inglis-Teller limit, for which the $(n-3)$ Stark manifold crosses the $n$S target state (the value of $E$ was deduced with an error of around $10\%$ from a measurement of the Stark map). Below the Inglis-Teller limit, both target state and ensemble lifetime agree with the theoretical predictions. From Fig. 2, two main effects are evident. Most strikingly, around the Inglis-Teller limit the observed target state lifetime drops by two orders of magnitude to a few  $\mathrm{\mu s}$ and remains low for larger $E$. The ensemble lifetime, by contrast, shows no particular variation at the Inglis-Teller limit but a distinct increase for larger field values. \\

While the increase of the ensemble lifetime above the Inglis-Teller limit can be explained by the fact that the central states of the $(n-3)$ Stark manifold with large angular momentum, and hence longer lifetime, are mixed into the target $nS$ state, at present we have no simple physical picture for the drastic behaviour of the target state lifetime. One possible explanation might be electric field inhomogeneities and state changes at the avoided crossings in the Stark manifold due to atomic motion, but thus far we have not been able to experimentally verify that hypothesis.\\

\begin{figure}[htbp]
\begin{center}
\includegraphics[width=9cm]{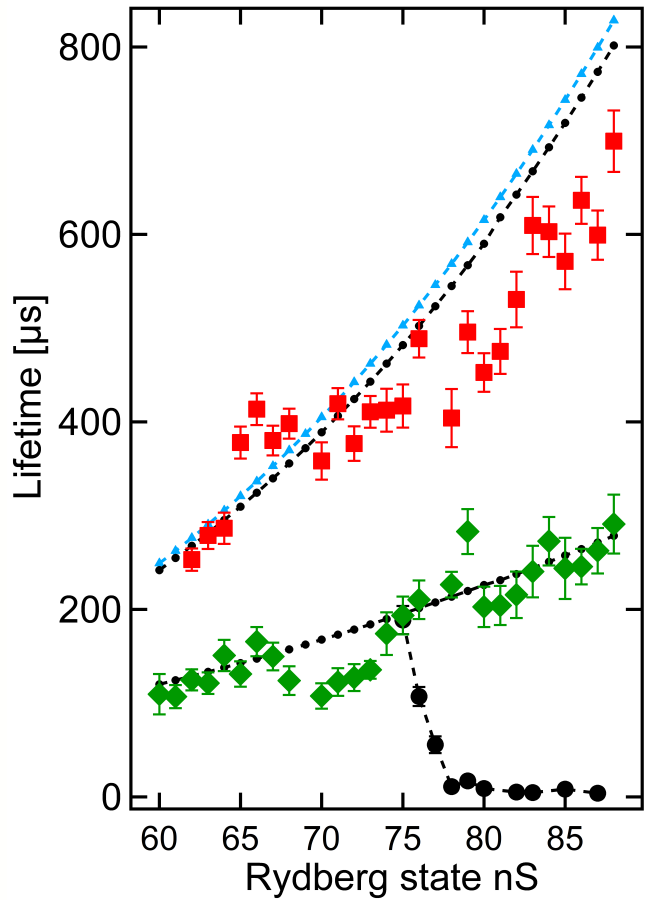}
\caption{Measurement of target state (green diamonds) and ensemble lifetimes (red squares) of high-lying Rb Rydberg states. The small black circles (connected by dashed lines to guide the eye) are the results of numerical simulations of the coupled black-body re-population and spontaneous decay processes. The blue circles represent the target state lifetime at zero temperature. The black circles (connected by a dashed line to guide the eye) are the target state lifetimes without compensation of the electric field in the cell (that field is around  $215\,\mathrm{mV/cm}$, corresponding to the Inglis-Teller limit for the 76S state; for the other measurements in this figure the residual electric field in the cell was less than $90\%$ of the Inglis-Teller limit.).  Note that for $n=60$ and $n=61$  no measurement for the ensemble lifetime is reported, as the lower limit of $n=60$ for reliable field ionization leads to systematic errors, but states below $n=60$ can be populated by black-body radiation.  Error bars are derived from the exponential fit. }
\label{default}
\end{center}
\end{figure}

From the above discussion, we conclude that in the absence of interaction effects between Rydberg atoms, our measurement protocol allows us faithfully to deduce the target state and ensemble lifetimes for Rydberg states with an Inglis-Teller limit greater than the $100\,\mathrm{mV/cm}$ residual electric field  in our cell (see inset of Fig. 2), and hence an upper limit of $n = 88$. In Fig. 3 we report the measured lifetimes \cite{comment} for states with $60\leq n\leq88$ (the lower limit is due to the upper limit of our ionization field). In order to compare our results with theoretical predictions, we used a numerical model taking into account both spontaneous decay as well as black-body induced population redistribution between the ensemble Rydberg states (taking into account states up to $\mathrm{L}=3$), including the possibility of a re-population of the target state from the support states. The latter process leads to an observed target state lifetime that is slightly longer (up to $10 \%$) than the values usually quoted in the literature, which only take into account the departure from the target state due to black-body induced transitions. We find good agreement between the numerical calculation and our measurements for the target states, with a slight exception around $n=72$ (which we will discuss below). In  Fig. 3 we also report the target state lifetimes measured without the electric field compensation; in that case, a sharp drop in the observed lifetime occurs around $n=76$, above which the Inglis-Teller limit is below the uncompensated background field. \\

Using our numerical model  we can also calculate the expected ensemble lifetimes, and find reasonable agreement with our experimental results. Obviously, the ensemble lifetimes are considerably longer (up to a factor of 3) than the target state lifetimes, as they are due to multi-step black-body redistribution processes to higher angular momentum states inside the ensemble (either with or without a change in $n$), along with spontaneous decay, for which the lifetimes become longer as the angular momentum increases. From those calculations we find that in our experiments starting from $n$S states, during the evolution of the system P, D and F states are significantly populated ({\em i.e}., more than a few percent of the total population), whereas higher angular momentum states can be neglected. Intuitively, right after the excitation of the target state, one expects the decay rate of the ensemble to be dominated by spontaneous decay of the target state to low-lying states, whereas all other processes leave the ensemble number unchanged. The ensemble lifetime should, therefore, approximate the Rydberg state lifetime at zero temperature. From Fig. 3 it is evident that the simulation for the ensemble lifetimes confirms this, and the experimental results also support that interpretation. \\

We now turn to the discrepancies between our experimental results and the numerical calculations. For the target state lifetimes, there is a localized deviation with lifetimes shorter than expected by up to $25\%$ around $n=72$, for which the transition frequencies to nearby $n$P and $(n-1)$P states are around $10\,\mathrm{GHz}$. One reason for that deviation could be the presence of microwave radiation at those frequencies (owing to the large dipole moment of those transitions, microwave powers in the tens of pW regime would be sufficient to cause the observed deviation). In preliminary experiments \cite{archimi19}, for which we added additional field electrodes to better compensate the background field, we were able to rule this out as a possible cause by shielding the apparatus with aluminium foil (using Autler-Townes splitting of an 91S state coupled to a 90P state induced by an external microwave source, we measured a shielding factor of around 4). \\

This leaves a variation in the black-body spectrum itself as a plausible explanation (other possible causes for enhanced decay, such as superradiance  \cite{gross79,wang07,zhou16,grimes17}, are unlikely to play a role in our system due to the small number of excitations involved and the narrow range of parameters for which such effects are expected to occur in multi-level systems  \cite{sutherland17}). In fact, it is well known that the assumptions made in the derivation of the Planck formula for black-body radiation are no longer valid when the wavelength of the radiation becomes comparable to the size of the black body, as the density of modes is strongly modified close to the longest-wavelength modes supported by the cavity \cite{garcia08,reiser13,fernandez18} (signatures of such an effect were detected experimentally in \cite{beterov99}). The dimensions of our vacuum cell (internal cross-section $1.8\,\mathrm{cm}$ x $2.4\,\mathrm{cm}$, with a thin coating of adsorbed Rb atoms on the inner walls) and of the surrounding support structures and coils are of order a few centimetres. The wavelengths corresponding to the transitions from nS states with $70\leq n\leq75$ to the closest $n^\prime$P states are between $2.5\,\mathrm{cm}$ and $3.3\,\mathrm{cm}$ and, therefore, comparable to those expected for the lowest modes inside the low-quality factor cavity formed by the glass cell. In addition, black-body radiation stemming from the surrounding support structures can also be modified compared to the Planck formula due to finite-size effects, further modifying the expected black-body spectral intensity seen by the Rydberg atoms.  \\

Regarding the ensemble, we find deviations of up to $30\%$, particularly towards shorter lifetimes for $n>70$, but also towards longer lifetimes. Again, these deviations might be partly due to the presence of microwave radiation with a spectral intensity that differs from the predictions of the Planck formula. Furthermore, as the ensemble lifetimes are several hundred microseconds, and hence measurements were taken up to several milliseconds, systematic effects due to the expansion and free fall of the atomic cloud might lead to an underestimation of the ensemble lifetimes. We have checked that the presence of the MOT lasers after $1500\,\mathrm{\mu s}$ (which are switched back on in order to minimize losses from the MOT during the experimental cycle, which is repeated at a frequency of $4\,\mathrm{Hz}$, and keep the number of atoms constant) does not influence the measurements. \\

In conclusion, we have presented a method for measuring the target state and ensemble lifetimes of high-lying Rydberg states based on a hybrid field ionization and optical de-excitation technique. The measured lifetimes are in good overall agreement with numerical calculations. Our method is suitable for measuring subtle deviations from the theoretically predicted lifetimes due to, for instance, finite-size modifications of the black-body spectrum or other effects such as super- or subradiance. In future experiments we plan to investigate the observed deviations more closely and to extend our measurement protocol to P, D and F states, which will give us further insight into possible modifications of the black-body spectrum. \\

We thank Thomas Gallagher, J\'{o}szef Fort\'{a}gh, Markus Greiner, David Petrosyan, Nikola \v{S}ibali\'{c}, Levi Sch\"{a}chter and Darrick Chang for discussions, and Alessandro Tredicucci and Giorgio Carelli for the loan of microwave equipment. This work was funded by the H2020-FETPROACT-2014 Grant No. 640378 (RYSQ). I.I.R. and I.I.B. were supported by the Russian Foundation for Basic Research under Grant No. 17-02-00987 (for lifetime calculations), by the Russian Science Foundation under Grant No. 18-12-00313 (for multi-step transitions induced by black-body radiation), and by Novosibirsk State University.



\bibliographystyle{apsrev4-1}

\bibliography{BibliographyVdW_V6} 

\begin{thebibliography}{10}
\bibitem{comment} For simplicity, we fit single exponential curves to our data in order to extract a single value for the lifetime, although the decay process is  not perfectly exponential. In practice, we find that such a fit is typically a reasonable approximation.
\end{thebibliography}

\end{document}